\documentclass[12pt]{article}

\include{epsf}
\include{psfig}
\input{psfig.sty}

\title{A subjective distance between stimuli: quantifying
the metric structure of representations}

\author{\normalsize D. Oliva$^1$, I. Samengo$^2$, S. Leutgeb$^3$ and
S. Mizumori$^4$}

\date{\small 1. Laboratorio de Neurobiolog\'{\i}a de la Memoria, Departamento
de Fisiolog\'{\i}a, Biolog\'{\i}a Molecular y Celular, Facultad de
Ciencias Exactas y Naturales, Universidad de Buenos Aires, Ciudad
Universitaria, 1428, Buenos Aires, Argentina. \\
2. Centro At\'omico Bariloche (8400) San Carlos de Bariloche,
R\'{\i}o Negro, Argentina. \\
3. Centre for the Biology of Memory, Norwegian University of
Science and Technology, NO-7489 Trondheim, Norway. \\
4. Psychology Department, University of Washington, Seattle, USA.}

\begin{document}

\maketitle

\begin{center}

{\bf Abstract}

\parbox{12cm}{As subjects perceive the sensory world, different
stimuli elicit a number of neural representations. Here, a
subjective distance between stimuli is defined, measuring the
degree of similarity between the underlying representations. As
an example, the subjective distance between different locations
in space is calculated from the activity of rodent's hippocampal
place cells, and lateral septal cells. Such a distance is
compared to the real distance, between locations. As the number
of sampled neurons increases, the subjective distance shows a
tendency to resemble the metrics of real space. }

\end{center}

\section{How different are two stimuli perceived?}

\label{distancias}

Consider a subject that is labeling the elements of a given set of
stimuli $S = \{s^1, s^2, ..., s^N\}$. Every time a stimulus $s^j
\in S$ is shown, he or she identifies it as $s^k \in S$, where
$j$ may or may not be equal to $k$. Successful trials are those
where the stimulus is correctly identified, that is, when $j = k$.
By writing down the succession of presented stimuli, and in each
case, the response of the subject, one can build up a list of
pairs $(s^j, s^k)$, where the first element, $s^j$, is the real
stimulus, and the second one, $s^k$, is the choice made by the
subject.

Notice that by looking into the table, one cannot determine
precisely what the subject actually perceived, since only the
final choice $s_k$ is accessible. The subject may, in fact,
hesitate to identify the stimulus as a member of $S$, or even
think that it does not truly match any of the $s^i$. However, even
if the mental representation elicited by stimulus $s^j$ is
unknown, the experimentalist can assess that under the
requirement to classify the stimulus as an element of $S$, the
subject chooses $s^k$. In a way, whatever the neural activity
brought about by $s^j$, out of all the elements in $S$, the one
whose representation is most similar to the actual one, is $s^k$.
The object of the present work is to provide a quantitative
measure of such a criterion of similarity. The approach does not
rely on a model of mental representations, it only makes use of
the statistics of actual and chosen stimuli.

\begin{figure}[htdf]
\begin{center}
\leavevmode \epsfysize=5truecm \epsfxsize=4truecm
\epsffile{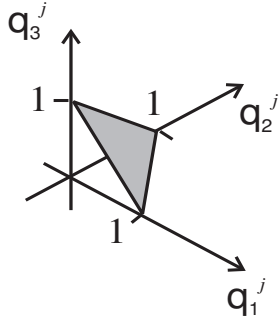}
\end{center}
\caption{Domain ${\cal D}$ where the vector ${\bf q}_j$ can
exist, when $N = 3$.} \label{f00}
\end{figure}

In what follows, we assume that a sufficiently large number of
samples has been taken, so that the conditional probability $Q(s^k
| s^j)$ of showing $s^j$ and perceiving $s^k$ may be evaluated,
for all $j$ and all $k$. The matrix $Q$ is henceforth called the
{\sl confusion matrix}. The elements of matrix $Q$ are positive
numbers, ranging from $0$ to $1$. In addition, normalization must
hold,
\begin{equation}
\sum_k Q(s^k | s^j) =1. \label{eio}
\end{equation}
It should be noticed that $Q$ need not be symmetric. For any
fixed $j$, one can define an $N$-dimensional vector ${\bf q}^j$
such that its $k$-th component is equal to $Q(s^k | s^j)$. The
positivity of the elements of $Q$ and the normalization condition
Eq. (\ref{eio}) determine a domain ${\cal D}$, where ${\bf q}^j$
can live. It is a finite portion of a hyperplane of dimension $N
- 1$. Figure \ref{f00} depicts the domain ${\cal D }$ for $N = 3$.

For some sets of stimuli, the confusion matrix may show
clustering. That is, choosing a convenient ordering of the
stimuli, $Q$ may show a block structure, where cross elements
between stimuli belonging to different blocks are always zero. In
this case, the stimuli belonging to different blocks are never
confounded with one another. Moreover, they are never confounded
with a third common stimulus. The phenomenon of clustering exposes
a very particular structure perceived by the subject in the set
of stimuli.

In this work, we are interested in studying the statistics of
mistakes. More specifically, if the subject happens to
systematically confound, for example, two of the stimuli, it could
be argued that from his or her subjective point of view, those two
stimuli are particularly similar. We would like to quantify such
an amount of similarity by introducing a distance between
stimuli. This distance, being defined with the statistics of
mistakes, is of course subjective.

It may happen that in a particular experiment, the subjective
distance between stimuli can actually be explained in terms of
some physical parameter qualifying the stimuli (orientation angle,
pitch, colour, etc). But it could also happen that the subjective
metric structure had no physical correlate in the stimuli
themselves, but instead depended on the previous semantic
knowledge of the subject, or on the presence or absence of
distractors, or on the statistical distribution with which the
stimuli are presented, or on the attention being paid by the
subject. The aim, hence, of defining a subjective distance
between stimuli, is to provide a quantitative measure that may
serve to determine the degree up to which these different
processes---if present---contribute to the confusion of some
stimuli with others, or, in contrast, to their clear
differentiation.

In order to define a distance between stimuli, it is necessary to
have a notion of equality. Here, two stimuli $i$ and $j$ are
considered subjectively equal if ${\bf q}^i = {\bf q}^j$. That is,
if for all $k$, $Q(s^k|s^i) = Q(s^k|s^j)$. Hence, in the notion
of equality not only the way stimulus $i$ is confounded with
stimulus $j$ is relevant. One must also compare the way each of
those two stimuli are confounded with the rest of the elements in
set $S$. If one of them is perceived as very similar to a third
stimulus $k$, but the other is not, then a noticeable difference
between $i$ and $j$ can be pointed out, and the stimuli cannot be
considered subjectively equal.

The equality of ${\bf q}^i$ and ${\bf q}^j$, in addition, is not
equivalent to a high confusion probability, between the two of
them. If stimulus $i$ is always perceived as stimulus $j$ and vice
versa, then---taking $s^i$ and $s^j$ as representing the first and
second components, respectively, in the ${\bf q}$ vectors---
$({\bf q}^i)^t = (0, 1, 0, ..., 0)$, whereas $({\bf q}^j)^t = (1,
0, 0, ..., 0)$. In this case, the two stimuli are perfectly
distinguishable from one another. The fact that the subject
chooses to label stimulus $i$ as $j$ (and vice versa) does not
mean he or she makes confusion between them. It is only a
question of names. Correspondingly, it may happen that two
stimuli are never confounded with one another, and yet they are
equal. This happens when $Q(s^i|s^j) = Q(s^j|s^i) = Q(s^i|s^i) =
Q(s^j|s^j) = 0$, and in addition, $Q(s^k|s^i) = Q(s^k | s^j)$,
for all $k$ different from $i$ and $j$.

Starting from the notion of subjective equality, in the next
section a number of desirable properties of a subjective distance
are discussed. Out of all the distances that fulfill  these
requirements, a single one is selected, in section 3. Next, in
section 4, the relationship of our subjective distance to other
measures of similarity is discussed. Section 5 extends the
definition of subjective distance to the case where the response
of the subject is given as a neural pattern of activity. In
section 6 an example is presented, using extracellular recordings
from the rodent hippocampus and lateral septum. Finally, in
section 7, a brief summary of the main ideas and results is given.

\section{Properties of a subjective distance}

What are the desirable properties of a subjective distance?
First, since the distance $D$ between elements $i$ and $j$ is
intended to reflect the statistics of confusions upon
presentation of these two stimuli, it is convenient to define it
in terms of the vectors ${\bf q}^i$ and ${\bf q}^j$. As a
distance, it is required to fulfill the following conditions:

\begin{enumerate}
\item $D({\bf q}^i, {\bf q}^j) \ge 0$, and
$D({\bf q}^i, {\bf q}^j) = 0 \Leftrightarrow {\bf q}^i = {\bf
q}^j$.

\item $D$ is symmetric: $D({\bf q}^i, {\bf q}^j) = D({\bf q}^j, {\bf q}^i)$.

\item $D$ obeys the triangle inequality: $D({\bf q}^i, {\bf q}^j) +
D({\bf q}^j, {\bf q}^k) \ge D({\bf q}^i, {\bf q}^k)$.
\end{enumerate}

These are general requirements, defining a distance. The third
condition implies that the set of ${\bf q}$ vectors that lie all
at the same distance of one particular ${\bf q}^i$ conform a
convex figure, in the domain ${\cal D}$.

In addition, in the present case, the distance between two
elements should not depend on the ordering of the stimuli. Hence,
if the components $k$ and $\ell$ are interchanged, in both ${\bf q
}^i$ and ${\bf q}^j$, the distance $D({\bf q}^i, {\bf q}^j)$
should remain invariant. That is, if $C^{k\ell}$ is a matrix that
interchanges the $k$-th and $\ell$-th component, then
\begin{enumerate}
\setcounter{enumi}{3}
\item $D({\bf q}^i, {\bf q}^j) = D(C^{k\ell}{\bf q}^i,
C^{k\ell}{\bf q}^j)$.
\end{enumerate}
This requirement, though plainly obvious from the intuitive point
of view, imposes quite serious restrictions. Consider, for
example, all the distances $D({\bf q}^i, {\bf q}^j)$ that can be
derived from a scalar product $<,>$, namely, $D({\bf q}^i, {\bf
q}^j) = \sqrt{<{\bf q}^i - {\bf q}^j,{\bf q}^i - {\bf q}^j>}$.
Once an orthonormal basis is given, this may be written as
$D({\bf q}^i, {\bf q}^j) = \sqrt{({\bf q}^i - {\bf q}^j)^t M
({\bf q}^i - {\bf q}^j)}$, where $M$ is any hermitian, positive
definite matrix representing the scalar product. Condition 4
imposes symmetry among the components of the vectors, which means
that $M$ must be proportional to the unit matrix. Therefore, out
of all the distances that have a scalar product associated to
them, the only one that fulfills condition 4 is the Euclidean
distance---apart from a scale factor, fixing the units.

What should be the meaning of the maximum subjective distance?
The maximum distance should be reserved to those pairs of objects
which the subject distinguishes unambiguously from one another:
that is, to those $s^i$ and $s^j$ that are never confounded with
a common stimulus. Mathematically, this means that for each $k$,
either $Q(s^k | s^i)$ or $Q(s^k | s^j)$ (or both) must vanish.
That is, whenever $Q(s^k | s^i) \ne 0$, $Q(s^k| s^j) = 0$ (and
vice versa). In this case, whatever the response of the subject to
stimulus $s^i$, it never coincides with his or her response to
stimulus $s^j$. This situation corresponds to the intuitive
notion of unambiguous segregation: the response of the subject to
stimulus $i$ is enough to ensure that the stimulus was {\sl not}
$j$. And vice versa, the response to stimulus $j$ is enough to
discard stimulus $i$.

The fifth requirement, hence, reads
\begin{enumerate}
\setcounter{enumi}{4}
\item if $D({\bf q}^i, {\bf q}^j)$ is maximal if and only if $s^i$
and $s^j$ are unambiguously segregated. And conversely, if $D({\bf
q}^i, {\bf q}^j)$ is not maximal, then $s^i$ and $s^j$ are not
unambiguously segregated.
\end{enumerate}

Imposing requirement 5 ensures that the stimuli that are
unambiguously segregated are all at the same distance, no matter
any other particular characteristic of the stimuli. And
conversely, if two stimuli do not elicit segregated responses,
they are not allowed to be at the maximum distance. Condition 5
establishes the cases that correspond to the maximum distance, in
the same way that condition 1 does to the minimum distance.
Adding the triangular inequality 3, ensures that those pairs of
stimuli whose distance lies in between of the minimal and the
maximal one, be consistently ordered.

Condition 5 has two important consequences. In the first place, it
ensures that the clustering structure present in $Q$ is also
reflected in the matrix of distances $D$. In $D$, of course, cross
terms between different blocks are not equal to zero, but to the
maximum distance. Conversely, if $D$ shows a block structure, it
may be shown that $Q$ has the same block structure. In the second
place, if the maximum distance is finite, a similarity matrix $S$
may be defined, $S = d_M I - D$, where $d_M$ is the maximum
distance, and $I$, the unit matrix. The matrix $S$ also inherits,
if present, the clustering structure of $Q$. This correspondence
between the clustering structure of $D$ (and of $S$) with the one
of $Q$ cannot be ensured, if condition 5 is not fulfilled.

The Euclidean distance $D_{\rm E}({\bf q}^i, {\bf q}^j) =
\sqrt{\sum_k (q^i_k - q^j_k)^2}$ does not fulfill condition 5.
Taking into account that the ${\bf q}$ vectors are normalized
(Eq. \ref{eio}), the maximum value of the Euclidean distance
between two stimuli is $\sqrt{2}$. It can be attained, for
example, for $({\bf q}^1)^t = (1, 0, 0, 0)$ and $({\bf q}^2)^t =
(0, 1, 0, 0)$. In this example, in fact, stimulus 1 shares no
common response with stimulus 2. However, not all stimuli with no
common responses lie at the maximum Euclidean distance. Consider,
for example, $({\bf q}^3)^t = (1/2, 1/2, 0, 0)$ and $({\bf q}^4)^t
= (0, 0, 1/2, 1/2)$. Though showing disjoint response sets, their
Euclidean distance is equal to 1, which is less than the maximum
distance. This means that none of the distances that can be
associated to a scalar product are useful, as a measure of
subjective dis-similarity.

\section{Choosing a subjective distance}
\label{d1}

There are still many distances fulfilling requirements 1 - 5. In
what follows, a single one is selected, on the basis of a maximum
likelihood decoding. Imagine that someone observing the subject's
responses to either stimulus $i$ or $j$ has to guess which of the
two has been presented. For the moment, for simplicity we assume
that both stimuli appear with the same frequency; this requirement
will be abandoned later on. We assume the observer is familiar
with the confusion matrix of the subject. There are several ways
in which he or she can decide between stimuli $i$ and $j$, given
the subject's response. Here, a maximum likelihood strategy is
considered, since this is the algorithm that maximizes the
fraction of stimuli correctly identified. It consists of taking
the choice of the subject - say, stimulus $k$- and deciding
whether the actual stimulus was $i$ or $j$ on the basis of which
of them has the largest ${\bf q}_k$ component. If $Q(s^k | s^i) >
Q(s^k | s^j)$, the observer chooses stimulus $i$, if the opposite
holds, the observer chooses stimulus $j$. If both conditional
probabilities are equal, then the observer chooses any of the two
stimuli, with equal probabilities.

Under this scheme, the fraction of times the observer correctly
identifies stimulus $i$ is
\begin{equation}
P(s^i|s^i) = \sum_{k / q_k^i > q_k^j} q_k^i + \frac{1}{2}\sum_{k
/ q_k^i = q_k^j} q_k^i.
\end{equation}
The fraction of times stimulus $j$ is presented, but the observer
chooses stimulus $i$ is
\begin{equation}
P(s^i|s^j) = \sum_{k / q_k^i > q_k^j} q_k^j + \frac{1}{2}\sum_{k
/ q_k^i = q_k^j} q_k^i.
\end{equation}
Correspondingly, when the observer decides for stimulus $j$
\begin{eqnarray}
P(s^j|s^j) &=& \sum_{k / q_k^j > q_k^i} q_k^j + \frac{1}{2}\sum_{k
/ q_k^i = q_k^j} q_k^j. \\
P(s^j|s^i) &=& \sum_{k / q_k^j > q_k^i} q_k^i + \frac{1}{2}\sum_{k
/ q_k^i = q_k^j} q_k^j.
\end{eqnarray}
The distance $D({\bf q}^i, {\bf q}^j)$ is defined as the
difference between the fraction of correct and incorrect maximum
likelihood choices, namely,
\begin{eqnarray}
D({\bf q}^i, {\bf q}^j) &=& \frac{1}{2} \left[P(s^i|s^i) -
P(s^i|s^j)\right] + \frac{1}{2} \left[P(s^j|s^j)- P(s^j|s^i)
\right] \nonumber \\
&=& \frac{1}{2}\sum_{k = 1}^N |q_k^i - q_k^j|. \label{definicion}
\end{eqnarray}
In other words, the distance between stimulus $i$ and stimulus
$j$ is defined in terms of the performance of the maximum
likelihood decoding, assuming that the response statistics of the
subject are known. This definition is easily shown to fulfills
all 1-5 conditions. A distance equal to zero means that the
observer is deciding at chance, between the two stimuli. Given
that he uses a maximum likelihood strategy, that means that the
two underlying vectors are equal. A distance equal to 1 implies
that the observer always makes the right choice.

\begin{figure}[htdf]
\begin{center}
\leavevmode \epsfysize=5truecm \epsfxsize=5truecm
\epsffile{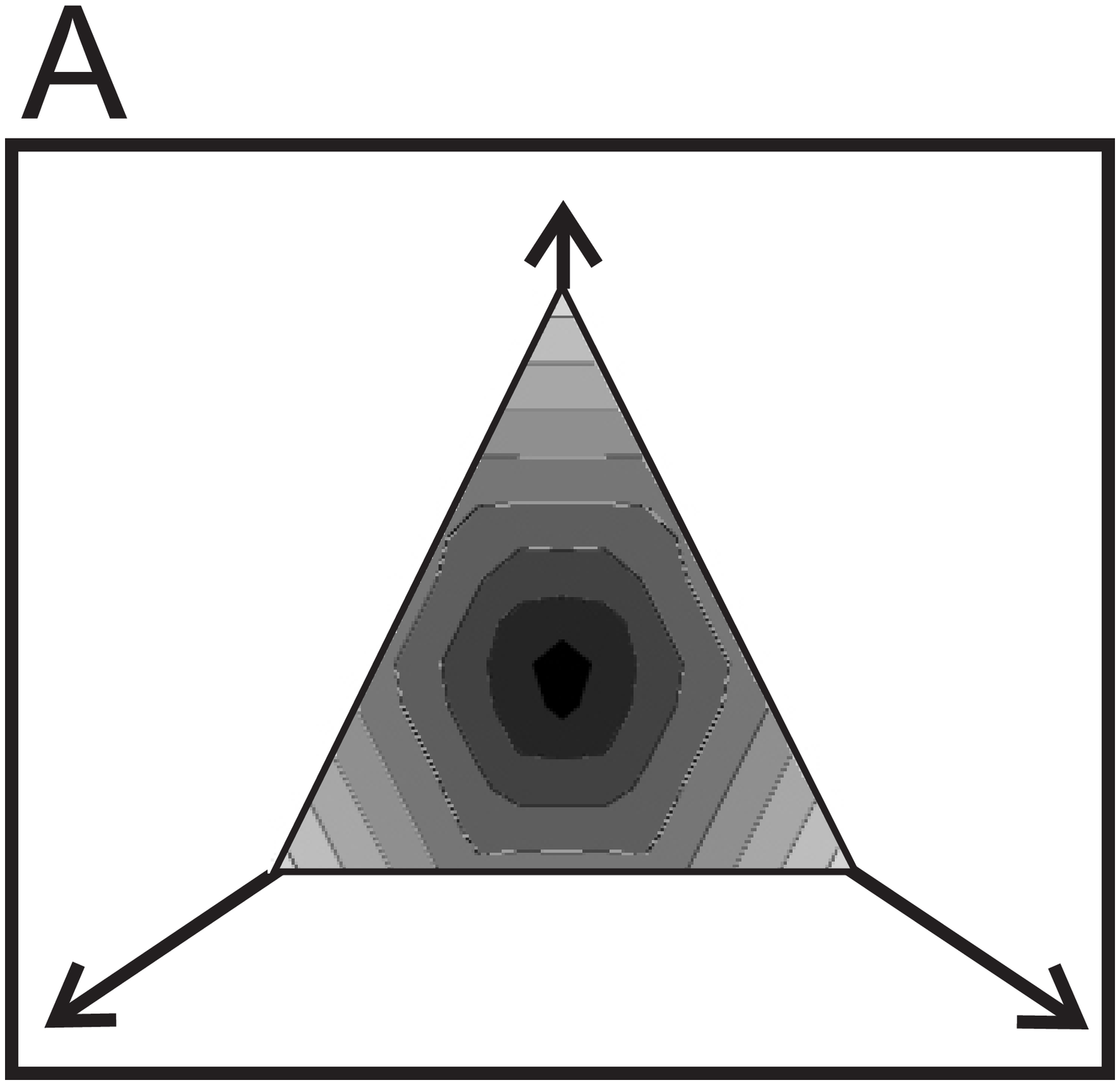} \epsfysize=5truecm \epsfxsize=5truecm
\epsffile{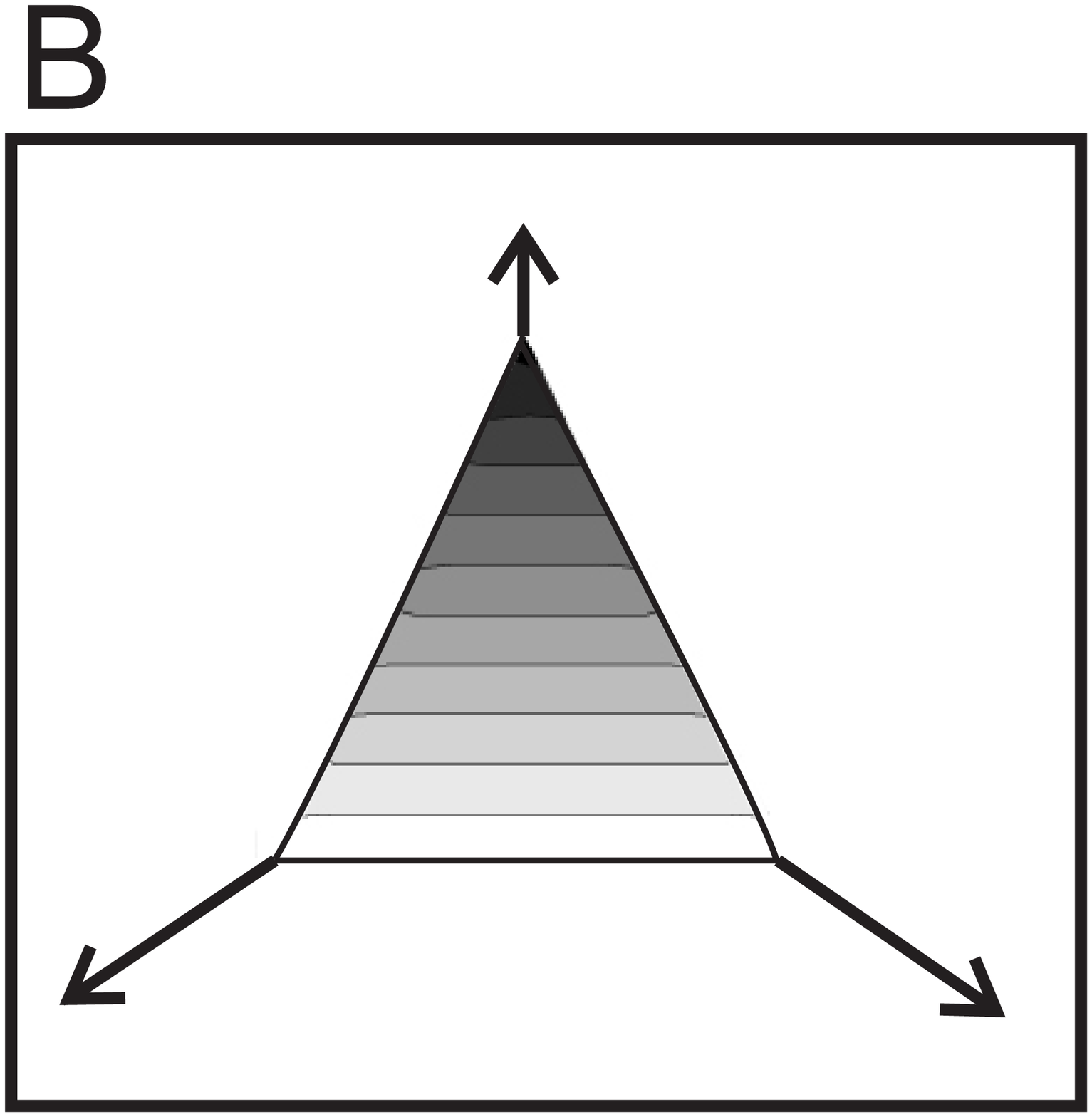}
\end{center}
\caption{Contour plot of the distance $D$ to a fixed vector ${\bf
q}^i$, in ${\cal D}$. (a) ${\bf q}^i = (1/3, 1/3, 1/3)$, and (b),
${\bf q}^i = (1, 0, 0)$}. \label{f01}
\end{figure}

In what follows, some mathematical properties of the distance $D$
are analyzed. In order to get a geometric flavor of $D$, figure
\ref{f01}
shows a contour plot of the distance of all the vectors in ${\cal
D}$ to the vectors $(1/3, 1/3, 1/3)$, in (A), and $(0, 0, 1)$, in
(B).

The subjective distance $D$ is translation invariant. That is, if
the vectors ${\bf q}^i$ and ${\bf q}^j$ are displaced by a fixed
vector $\Delta{\bf q}$, the distance between them remains
unchanged. Mathematically,
\begin{equation}
D({\bf q}^i, {\bf q}^j) = D({\bf q}^i + \Delta{\bf q}, {\bf q}^j
+ \Delta{\bf q}). \label{h1}
\end{equation}
Equation (\ref{h1}) is valid for any displacement $\Delta {\bf
q}$. However, in the present context, the displacement should be
such that ${\bf q}^i + \Delta {\bf q}$ and ${\bf q}^j + \Delta
{\bf q}$ fall both inside the domain ${\cal D}$. This means that
the components of $\Delta {\bf q}$ must sum up to zero, and its
magnitude must be bounded (the value of the bound depends on the
location of ${\bf q}^i$ and ${\bf q}^j$).

As a further characterization, the distance $D$ between a
stimulus that is perfectly identified by the subject - say,
stimulus $i$ - and another stimulus $j$, is given. We take $({\bf
q}^i)^t = (1, 0, 0, ..., 0)$. In this case, $D({\bf q}^i, {\bf
q}^j) = \sum_{k = 2}^{N} q_k^j = 1 - q_1^j$. The distance between
${\bf q}^i$ and ${\bf q}^j$ is fully determined by $q_1^j$, it
does not matter whether the probability of {\sl not} selecting
stimulus 1 is spread out over the last $N - 1$ components of
${\bf q}^j$, or is entirely concentrated in a single one. This
result generalizes to any vector ${\bf q}^i$ having two or more
null components: the distance depends on the sum of those same
components of ${\bf q}^j$, and not on their individual values.

Finally, consider the case where the subject has a probability
$\alpha$ of identifying any of the stimuli correctly ($Q(s^i|s^i)
= \alpha$), and that whenever he or she makes a mistake, the error
is equally distributed among all other stimuli ($Q(s^i|s^j) =
\beta$, for all $i \ne j$). The normalization condition Eq.
(\ref{eio}) implies $\alpha + (N - 1)\beta = 1$. In this case,
$D(s^i, s^j) = |\alpha - \beta|$.

\subsection{Extension to continuous stimuli}

Consider the case where there is a continuous parameter $x$
labeling the stimuli, such that stimulus $i$ corresponds to an
interval of $x$ values ranging from $i \Delta$ to $i \Delta +
1/\Delta$, with $\Delta = 1 / N$. It is now convenient to vary $i$
between 0 and $N - 1$. The scale of $x$ is chosen in such a way
that its maximum value is 1. For large $N$, the confusion matrix
can be written as $Q(s^j|s^i) = u(j \Delta | s^i) \Delta$, where
$u(x|s^i)$ is a piecewise continuous probability density. The
distance between stimuli $i$ and $j$ reads
\begin{equation}
D({\bf q}^i, {\bf q}^j) = \frac{1}{2} \sum_k \left| u(k \Delta |
s^i) - u(k \Delta | s^j) \right| \Delta \rightarrow \frac{1}{2}
\int_0^1 |u(x | s^i) - u(x | s^j)| \ dx, \label{cont}
\end{equation}
where the right hand limit corresponds to making the number of
stimuli $N$ tend to infinity. In other words, the distance
between stimuli $i$ and $j$ is equal to the area between the two
corresponding densities. The normalization condition for $u(x|s)$
ensures that $D$ lies between 0 and 1. One can easily show that
its value remains invariant, when a different parametrization of
the variable $x$ is used---as long as the new variable is in a
one to one relation to $x$.

\subsection{Extension to stimuli with non uniform prior probabilities}

One could ask whether the measure $D$ can be extended to stimuli
which are not all presented with the same probability. Let
$Q(s^i)$ denote the probability of presenting stimulus $s^i$. In
this case, the maximum likelihood algorithm has to be substituted
by a maximum a posteriori one. That is, the observer chooses
stimulus $s^i$ upon response $s^k$ from the subject, whenever
$p_k^i = Q(s^k | s^i)Q(s^i)$ is larger than $p_k^j = Q(s^k |
s^j)Q(s^j)$, and vice versa. Whenever the $p_k^i = p_k^j$, the
choice between $s^i$ and $s^j$ is proportional to their
corresponding priors. In this case, the difference between the
correct and incorrect fractions of maximum likelihood estimations
is
\begin{equation}
D_0({\bf q}^i, {\bf q}^j) = \frac{1}{Q(s^j) + Q(s^i)} \sum_k
\left|p_k^i - p_k^j \right|. \label{nonun}
\end{equation}
This is, of course, also a valid distance between stimuli, since
it fulfills requirements 1 - 5. Its maximum value is also 1, and
it still carries the same block structure as $Q$.  Notice that in
this case, the distance not only depends on the subject's
perception---characterized by $Q(s^j|s^i)$---but also on the
statistics of the stimuli (described by $Q(s^i)$).

\section{Comparison with other measures of dis-si\-mi\-la\-ri\-ty}

The distance $D_0$ is no more than a geometrical view of the
matrix $Q(s^i, s^j)$. It has the advantage of being true distance,
that is, of obeying conditions 1 - 3, of section
\ref{distancias}. In addition, it fulfills the symmetry
constrains imposed by condition 4, and preserves the block
structure in $Q$, as ensured by condition 5. However, there are
also other distances that still obey requirements 1-5, the angle
between the vectors ${\bf q}_i$ and ${\bf q}_j$ can be taken as
an example. The advantage of $D_0$ is that it has a simple
interpretation in terms of maximum likelihood decoding
performance.

There have been previous attempts aiming at quantifying how
different two stimuli are perceived. Maybe the most similar to
ours was proposed by Green and Swets (1966), in their definition
of the discriminability $d'$ between two stimuli. Strictly
speaking, their approach can only be used when the response to
different stimuli is described by Gaussian functions whose mean
depends on the stimulus, but whose variance remains fixed. They
defined the discriminability $d'$ between two stimuli as the ratio
between the difference of the two corresponding mean values to the
standard deviation.

Can the concept of $d'$ be extended to more general response
distributions? Still in the Gaussian case, one can make a
correspondence between a given $d'$ value, and the expected
fraction of errors when estimating the stimulus from the response
of the subject, using a maximum likelihood decoding. In terms of
this correspondence, $d'$ not only represents the distance at
which the Gaussian functions sit from one another, but more
generally, the fraction of decoding mistakes---which is something
that does not depend on the shape of the probability distribution
of the responses of the subject. Large discriminability is
associated to a small probability of making a mistake. The scale
of the measure, however, since inherited from the Gaussian case,
has a non linear relationship with the fraction of mistakes.

With this idea in mind, $d'$ can be extended to non Gaussian
stimuli (Rieke, Warland, de Ruyter van Steveninck, and Bialek,
1997). Whatever the shape of $P(s^k | s^i)$ and $P(s^k | s^j)$,
given the response $s^k$, an observer can decide in favor of
stimulus $s^i$ or $s^j$ depending on which of them has a highest
probability of eliciting response $s^k$. For each pair of stimuli
$s^i$ and $s^j$ there will be, on average, a certain fraction
$f_e$ of errors. One could extend the definition of
discriminability to be the $d'$ value that would give, in the
Gaussian case, the same fraction $f_e$ of errors. This extension,
being intimately related to the performance of a maximum
likelihood decoding, is grounded on the same rationale as our
subjective distance $D$. The problem is that if one is
constrained to choose the $d'$ scale as to match the equivalent
Gaussian case, then the definition of discriminability does not
obey the triangular inequality. It is easy to construct an
example where $P(s^k | s^i)$ overlaps in certain region with
$P(s^k| s^j)$, which in turn, overlaps in a different region with
$P(s^k | s^\ell)$, but such that $P(s^k | s^i)$ and $P(s^k |
s^\ell)$ are never simultaneously different from zero. In this
case, $d'(s^i, s^j)$ and $d'(s^j, s^\ell)$ are both finite,
whereas $d'(s^i, s^j)$ is infinite. Hence, though both $D$ and
$d'$ can be defined in terms of maximum likelihood performance,
$D$ is a proper distance, whereas $d'$ is not.

Another well known notion of distance can be defined (for
continuous stimuli) in terms of the Fisher information metric
tensor $J$. There is a natural scalar product associated with
$J$, and also a notion of distance in the space of stimuli (see
Amari 1999). However, the entire Fisher geometry becomes
meaningless for discrete stimuli. Our aim is to show that even in
the discrete case, a definition of distance is possible.

The distance defined in terms of the Fisher metric tensor is a
bilinear form. Its matrix elements are not constant, but depend on
the point ${\bf q}$ that one is interested in. Hence, distances
are defined in terms of a curvilinear integral---which may
actually involve very difficult calculations. When two stimuli
have disjoint response sets, the Fisher distance between them
diverges. However, a Fisher distance between two stimuli equal to
infinity does not necessarily imply that the responses to those
two stimuli conform disjoint sets. The Fisher distance between
two stimuli may diverge, for example, when the probability
density $u(x | s)$ is discontinuous. This implies a discrepancy
with requirement 5.

The Kullback-Leibler divergence (Cover and Thomas, 1991) between
the vectors ${\bf q}^i$ and ${\bf q}^j$ can also be used to
measure how different two stimuli are perceived. It has the
appealing property of being intimately related to many concepts in
information theory, and as such, it has an information based
intuitive interpretation: it is a measure (in number of
additional bits of the mean code length) of the inefficiency of
assuming that the distribution of a given variable is ${\bf q}^i$
when its true distribution is ${\bf q}^j$. It is not a distance,
however, since it does not fulfill requirements 2 - 3. Its
symmetrized version is sometimes called the Jensen-Shannon
measure, which is still not a distance, because it does not obey
the triangular inequality 3. For this measure, condition 4 is
always true. The maximum value of the Jensen-Shannon divergence
is infinite. This value, however, is not only reached for pairs
of stimuli whose response sets are disjoint, but also whenever
there is a component $k$ such that $q_k^i = 0$, and $q_k^j \ne
0$. This means requirement 5 is not, in general, fulfilled.

There has been another previous proposal of a pseudo-distance
(Treves, 1997), which was also defined in terms of the confusion
matrix. As opposed to $D_0$, and also to the Jensen-Shannon
measure, the distance between stimuli $s^i$ and $s^j$ only depends
on $Q(s^i|s^j)$, $Q(s^j|s^i)$, $Q(s^i|s^i)$ and $Q(s^j|s^j)$,
(other stimuli do not appear). Just as the Jensen-Shannon
divergence, however, it does not fulfill requirements 3 and 5.

We claim that our measure allows to have a geometrical picture of
a set of stimuli. Multidimensional scaling (Young and Ham 1994,
Cox and Cox 2000) also aims at this goal. The two approaches,
however, bear certain differences, which are now discussed. The
use of $D_0$ aims at a definition of a distance. It is
particularly useful when one can vary a certain parameter in the
experiment (in the example below, the number of cells being
sampled) and wants to know the effect of that parameter in the
structure of confusions. Nevertheless, $D_0$ in itself does not
provide a way of visualizing the stimuli in a particular space.
Since the structure of confusions may depend on a very large
number of factors (as for example, the semantic knowledge of the
subject), there may be, actually, no small dimensional space
where the stimuli can be placed.

Multidimensional scaling, in contrast, starts with a given matrix
of distances, or dis-similarities. The algorithm is designed to
place the stimuli in a finite dimensional space producing the
minimum possible distortion of the pairwise distances. Except for
very particular sets of stimuli, it is an approximate method.
What is gained is the optimal set of coordinates of the stimuli,
out of the matrix of distances. The subjective distance could be,
in certain applications, a good starting point with which to feed
the multidimensional scaling algorithm.

Finally, we point out that our definition of subjective distance
makes special emphasis in the probability of confounding the
stimuli, as opposed to other possible measures, that stress the
differences in the elicited neural representations. In this
sense, our approach is, broadly speaking, complementary to Victor
and Purpura's (1997) proposal of constructing a metric in the set
of responses. There, different distances between spike trains were
considered. Each proposed distance captured specific aspects of
the neural response. For example, the distance between two spike
trains was either defined in terms of how different the timing of
individual spikes were, or how different the inter-spike
intervals were, and so forth. Their aim was to decide which of
those distances could better cluster the neural responses
corresponding to the different stimuli in their set, and in such
a way, to make inferences about the way stimuli were encoded into
spike trains. In the back of this reasoning is the assumption
that the stimuli themselves are all sufficiently different from
one another. Here, in contrast, we are interested in the perceived
distances and similarities of the stimuli, as can be deduced from
the structure of confusions.

\section{From neural representations to the subjective distance}

In order to use the distances defined in section \ref{d1}, the
conditional probabilities $Q(s^i | s^j)$ are needed. Such
probabilities may be extracted, as described in section
\ref{distancias}, from an experiment where the subject is asked
to identify the stimuli. However, this is not the only way a
matrix $Q$ can be obtained. In the case of non human subjects,
many times the stimuli are presented while the activity of one or
several neurons is being recorded by microelectrodes implanted
into the animal's brain. From these experiments, the conditional
probability $\rho(r^k | s^j)$ of recording response $r^k$ upon
presentation of stimulus $s^j$ may be calculated.

One way of deriving $Q$ from $\rho$ is to use a decoding
procedure. That is, to define a mapping going from the set of
neural responses to the set of stimuli (Rolls and Treves, 1998;
Dayan and Abbot, 2001; Rieke, Warland, de Ruyter van Steveninck,
and Bialek, 1997). Different rules defining such a transformation
give rise to different decoding procedures. Among them, one can
point out the maximum a posteriori decoding, where $r^k$ is
associated to the stimulus which maximizes $\rho(s^j|r^k) =
\rho(r^k | s^j) P(s^j)/\sum_\ell \rho(r^k | s^\ell) P(s^\ell)$,
that  is, the conditional probability of having presented
stimulus $s^j$ when response $r^k$ was measured. This rule is,
among all possible decoding rules, the one that maximizes the
fraction of correct decodings.

Another widely used option is the Bayesian approach, where
\begin{equation}
Q(s^j | s^i) = \sum_k \rho(s^j|r^k)\rho(r^k|s^i). \label{bayes}
\end{equation}
Strictly speaking, in this case there is no decoding, since one
does not choose a single stimulus for each response. One rather
keeps all the probability distribution for each stimulus, given
the neural response. The only assumption in the back of Eq.
(\ref{bayes}) is that the decoded stimulus is conditionally
independent of the actual stimulus, that is $P(s^j|r^k, s^i) =
P(s^j|r^k)P(r^k|s^i)$.

Passing from matrix $\rho$ to matrix $Q$ always means a loss of
information. The amount of information that is lost can sometimes
be estimated (Samengo, 2001). A maximum a posteriori decoding
maximizes the fraction of correct identifications (that is, the
trace $f$ of the resulting $Q$) but probably looses some of the
structure of $\rho$, whenever the response $r^j$ is not the most
probable response elicited by a given stimulus. A procedure like
Eq. (\ref{bayes}) is intended to preserve the statistical
regularities of mistakes, but typically implies a lower fraction
of correct choices $f$. Therefore, whenever neural responses are
used to construct matrix $Q$, one must bear in mind that the
results depend, at least partially, on the way of calculating $Q$.

The dependence of the $Q$ matrix on the decoding may seem perhaps
dangerous. Could one define a distance in terms of $\rho$, without
having to pass through matrix $Q$? Of course this is possible, in
general terms. One option, for example, would be to calculate the
mean response ${\bf m}^i$ to stimulus $i$, and then define the
distance between stimuli $i$ and $j$ as the distance $D_n$ between
the corresponding means\footnote{Of course, even more
sophisticated distances are possible, for example, one taking
into account the covariance matrix, or even higher moments of the
distributions $\rho(r^k|s^i)$. Here, we only use a very simple
definition of a distance $D_n$ in the response space, since we
only want to point out that there is a qualitative difference
between defining a distance in terms of the $Q$ matrix and in
terms of the $\rho$ matrix.}. This is a distance between vectors
living in the response set, not in ${\cal D}$. The sub-index $n$,
hence, stands for {\sl neural}. The distance $D_n$ may be a
sensible approach when one is interested in the neural
representations of the stimuli. It is not very convenient,
however, if the distance is meant to reflect the structure of
confusions. As is shown below, $D_n$ does not, in general,
fulfill condition 5.

In the entire subject, confusions are defined at the behavioral
level: the subject is asked to identify the stimuli. At the
neural level, one can only talk about confusion between two
stimuli when a decoding procedure is introduced. In this sense,
the process of decoding should not be viewed as an arbitrary
step: one shifts the question of how distinguishable two stimuli
are, to the problem of inferring the stimulus from the neural
response.

To test whether a distance defined in the response set fulfills
condition 5, we point out that for any reasonable decoding
procedure, two stimuli that elicit responses occupying convex,
non overlapping, portions of the response space are never
confounded with one another. Is this enough to have them at the
maximum $D_n$? In general, no. Being $D_n$ defined in terms of a
distance in the response set, the more separate the responses to
the two stimuli, the larger $D_n$. Hence, neural based distances
do not, in general, fulfill condition 5.

\section{Application to place cells: comparing the actual and the
subjective distance between two locations in space}

\label{placecells} It is known that many of the pyramidal cells
in the rodent hippocampus selectively fire when the animal is in
a particular location of its environment (O'Keefe and Dostrovsky,
1971). Such a response profile allows one to define, for each
cell, a {\it place field}, that is, the region in space for which
the neuron is responsive. A classical experiment in the study of
the rat's hippocampal neurophysiology consists in letting the
awake animal wander in a given environment, while the activity of
one or several of its hippocampal pyramidal cells is being
recorded (for an overview see, for example, O'Keefe (1971), or
Redish (2000)). In the experiment we analyze next, cells in the
lateral septum have also been recorded. The lateral septum
receives a massive projection from hippocampal pyramidal neurons,
the activity of septal cells has also been shown to be
informative about the location of the animal, though not as neat
as in hippocampal neurons.

The set of stimuli, in this case, consists of all the possible
locations where the animal can be placed. This set is endowed
with a natural metric: we all know what the distance between two
locations is. Other sets of stimuli, for example a set of pictures
of human faces, lack an obvious, natural distance. In order to
test our definition of subjective distance, it is convenient to
work with a metric set of stimuli, which allows the comparison of
$D_0$ with the natural {\sl physical} distance in the set.

\begin{figure}[htdf]
\begin{center}
%\leavevmode \epsfysize=6truecm \epsfxsize=4.5truecm
%\epsffile{fig3ax.eps} \leavevmode \epsfysize=6truecm
%\epsfxsize=4.5truecm \epsffile{fig3bx.EPS}
\leavevmode \epsfysize=6truecm \epsfxsize=12truecm
\epsffile{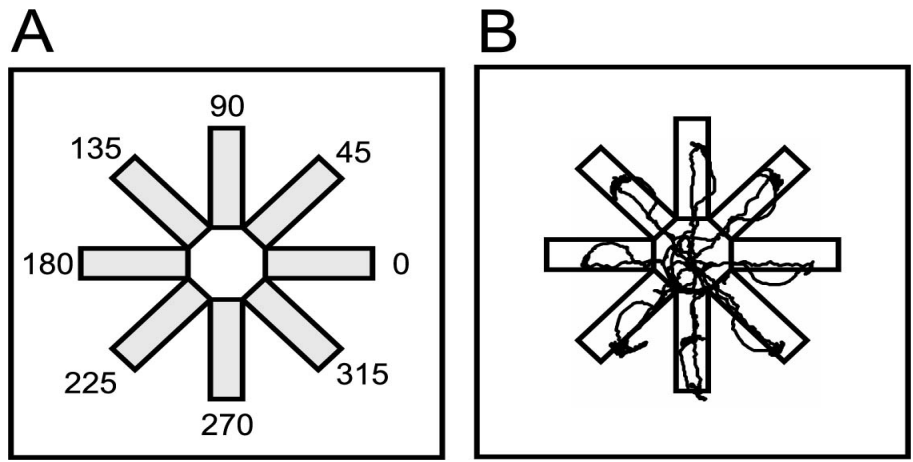}
\end{center}
\caption{(A) Eight arm maze where the animals move. (B) Path
traveled by a rat, in a given trial, as measured by the diode on
its head.  The rat goes straight to the end of the arm, and
drinks the chocolate milk. When turning round, it typically
sweeps its head in a circular movement, peeping outside the
maze.} \label{f1}
\end{figure}

A description of the experiment follows.  Nine young adult
Long-Evans rats were tested while moving in an 8 arm radial maze,
as shown in figure \ref{f1} (A).

Each arm contained a small amount of chocolate milk in its distal
part (for details see Leutgeb and Mizumori, (2000)). In a given
trial, a rat initially placed in the center of the maze, visited
the eight arms in a random order (see the example of figure
\ref{f1} (B)) taking the food reward. Several trials were
recorded per animal, sometimes in darkness and sometimes in light
conditions. Septal and hippocampal cells were simultaneously
registered while the animal moved on the maze. Single units were
separated using an on-line and off-line separation software.
Units were then classified according to their anatomical location
and the characteristics of the spikes. Hippocampal pyramidal
cells and lateral septal cells were identified.  In what follows,
each arm of the maze is taken as a different stimulus, and it is
labeled by its angle, as in figure \ref{f1} (A). The center of
the maze was excluded, so as to have eight similar, and evenly
visited stimuli. Thus, in the present experiment the physical
distance between any two stimuli is the (actual) angle between
the corresponding arms. The subjective distance is deduced from
the responses.

\begin{figure}[htdf]
\begin{center}
%\leavevmode \epsfysize=6truecm \epsfxsize=5truecm
%\epsffile{fig4ax.eps} \leavevmode \epsfysize=6truecm
%\epsfxsize=5truecm \epsffile{fig4bx.eps}
\leavevmode \epsfysize=6truecm \epsfxsize=10truecm
\epsffile{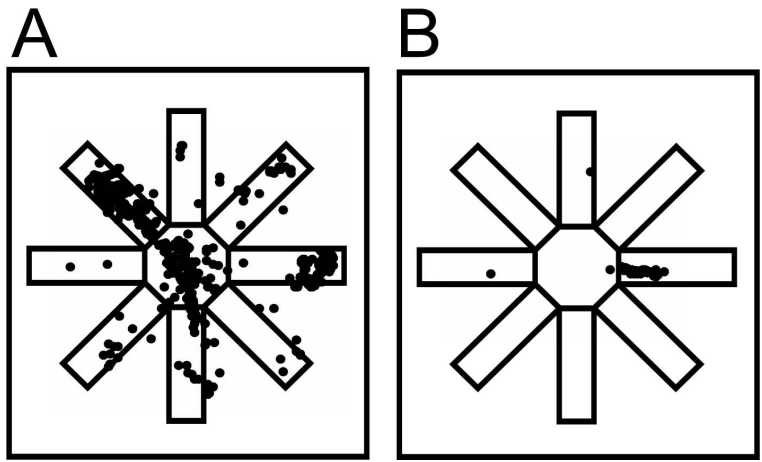}
\end{center}
\caption{Location of the animal when (A) a given lateral septal
and (B) a hippocampal pyramidal neuron fired a spike. The dots
that fall outside the maze indicate that the nose of the animal
is peeping outside the walls of the labyrinth. } \label{f2}
\end{figure}

As an example, in figure \ref{f2} the location of the animal is
shown, whenever (A) a lateral septal, and (B) a hippocampal place
cell fire a spike. It is clear that both cells fire selectively
when the animal is in a particular location on the maze, the
septal cell having a somewhat more distributed response.

The decoding procedure and the calculation of the $Q$ matrix was
carried out first for a single neuron, and then for pairs,
triplets and sets of 4 cells measured simultaneously. The
response of the animal was, correspondingly, a scalar, or a 2, 3
and 4 dimensional vector ${\bf r}$, where the component $r_c$
stands for the firing rate of cell $c$. Hence, upon entrance to
arm $s^j$, $r_c$ was calculated as the ratio of the number of
spikes fired by cell $c$ to the time spent in the arm $s^j$, in
that particular trial. To decode the arm corresponding to a given
response ${\bf r}$, the $M$ responses most similar to ${\bf r}$
were taken into account. Those $M$ firing rates, included $M_0$
that corresponded to the animal in arm $0^\circ$, $M_{45}$ to the
animal in arm $45^\circ$, and so forth. That is, $M = \sum_i
M_i$, and the response $r$ of the present trial is not included.
The probability $\rho(s^j|r)$ for the rat to be in arm $s^j$ when
response $r$ was observed was set as $M_{j} / M$. The decoded arm
was the one maximizing $\rho(s^j | r)$. If there was a draw
between two arms, the decoded one was chosen at chance between
those two, with probabilities that are proportional to the two
corresponding priors. With this procedure, $Q(s^i|s^j)$ is
defined as the fraction of times $s^i$ was decoded, whenever
$s^j$ was the actual arm.

The procedure was carried out for several values of $M$, ranging
from 2 to 20. Each $M$ gave rise to a different $Q$ matrix. We
observed that the fraction of correct decodings (the trace of $Q$)
typically showed a maximum, as a function of $M$. The $M$ for
which ${\rm tr}(Q)$ was maximal was taken to be the final one.
With this procedure, we obtained a $Q$ matrix for each neuron,
for each pair of neurons, for each triplet, and for each
quadruplet. In what follows, the results for the subjective
distance as derived from the chosen $Q$ matrices are shown.

\begin{figure}[htdf]
\begin{center}
\leavevmode \epsfysize=5truecm \epsfxsize=8truecm
\epsffile{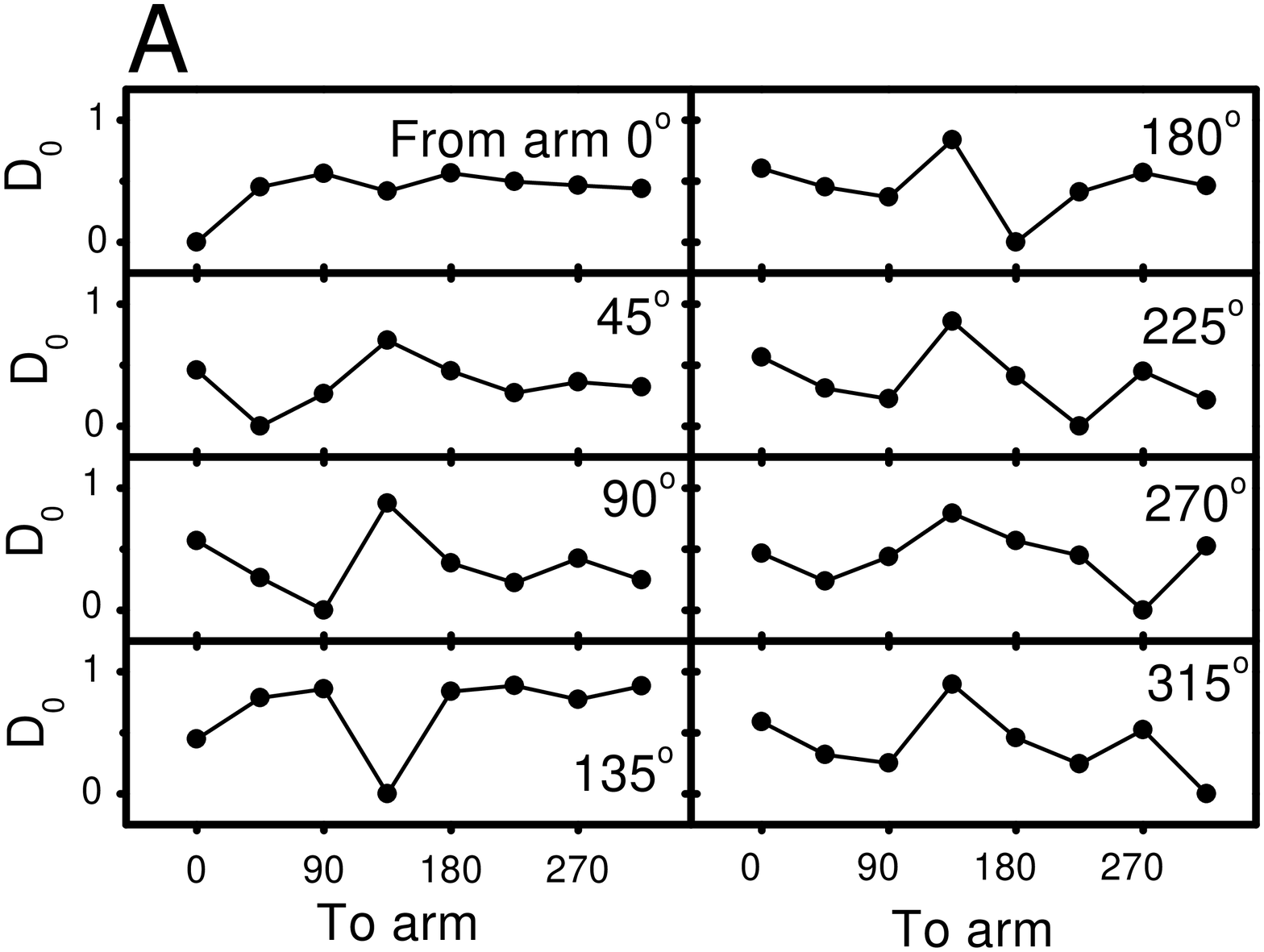} \leavevmode \epsfysize=5truecm
\epsfxsize=8truecm \epsffile{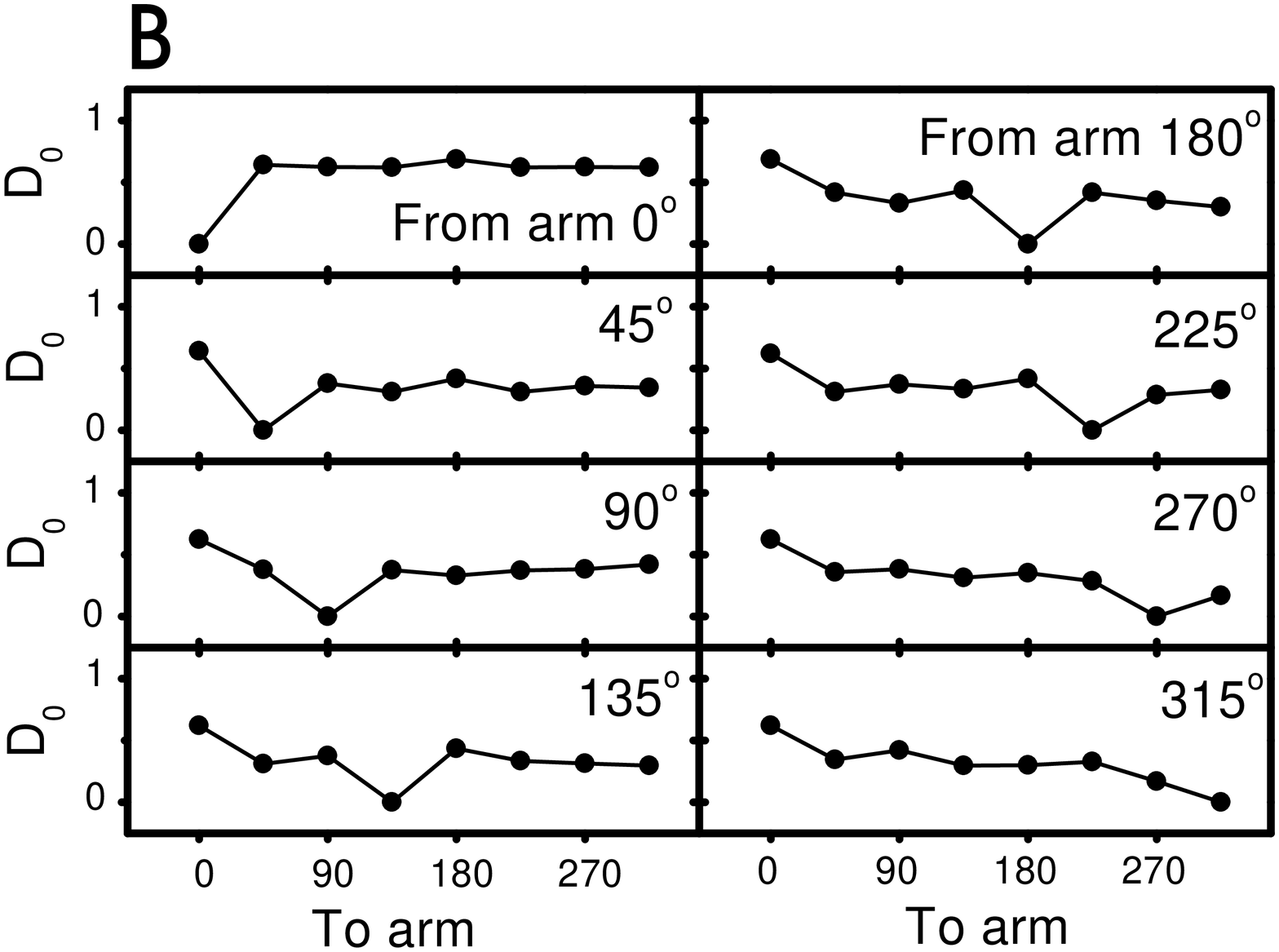}
\end{center}
\caption{Matrix of distances $D_0$ for (A) the lateral septal
cell of figure \ref{f2} (A), and (B) the hippocampal pyramidal
cell of figure \ref{f2} (B).} \label{f3}
\end{figure}

As examples from the single neuron behavior we show, in figure
\ref{f3},
the matrix $D_0({\bf q}^i, {\bf q}^j)$, for the same
cells of figure \ref{f2}. In each plot, the arm $i$ is kept fixed,
while the arm $j$ is varied (it does not really matter which is
$i$ and which is $j$, since the distance is symmetric). In all
graphs $D_0 = 0$ corresponds to $i = j$.

The plot in (A) corresponds to the lateral septum. It may be seen
that most arms are far away from the one at 135$^\circ$. Not
surprisingly, this is precisely the arm where the cell fires
most. If a strong, reliable response is obtained for a given
stimulus, and if this response differs from the response to all
other stimuli, then there is little probability to miss-identify
it. Among the remaining arms, the one closest to the one at
135$^\circ$ is the one at 0$^\circ$, and it corresponds to the
second largest firing rate.

\begin{figure}[htdf]
\begin{center}
\leavevmode \epsfysize=6truecm \epsfxsize=9truecm
\epsffile{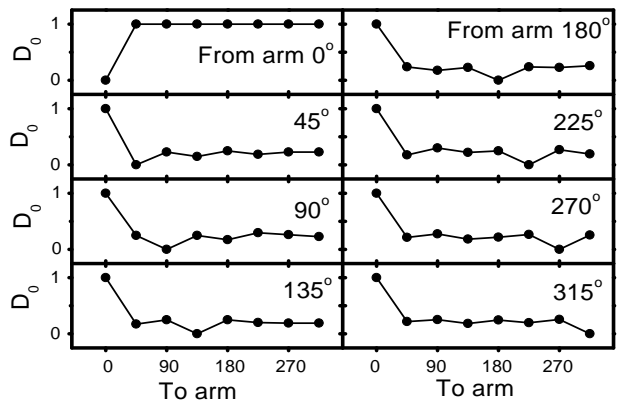}
\end{center}
\caption{Matrix of distances $D_0$ for the cell of figure
\ref{f2} (B), when all the trials with silent entrances to the
arm at 0$^\circ$ have been removed.} \label{f3a}
\end{figure}

In the case of the hippocampal cell (B), the arm that is most far
away to all others is the one at 0$^\circ$, the one where the
place field is located. Surprisingly, however, its distinction
from all other arms is not as clear as in the septal cell of part
(A), even though figure \ref{f2} (B) indicates that this cell is
much more selective to the location of the animal. The fact is
that figure \ref{f2} alone is not enough to depict the selectivity
of the cell, because it does not show the statistics. The rat in
(B) entered 18 times into the arm at 0$^\circ$, but in only half
of those trials the cell fired a burst of spikes. The other half
had no response. Therefore, a burst of spikes most probably means
the animal is the arm at 0$^\circ$, but a silent response does not
discard this arm.  In all those trials when there was no activity
upon entrance to the arm at 0$^\circ$, this arm can well be
confounded with any other arm. If all the silent entrances to the
arm at 0$^\circ$ are discarded, then the matrix of distances
changes drastically, as is shown in figure \ref{f3a}.
There, the distance from the arm at 0$^\circ$ to any other arm is
shown to be equal to one, implying no mistakes at all. This shows
that the trial to trial variability has an important influence on
the distance between any two stimuli, even for those stimuli that
may elicit very strong responses.

Every cell has a different spatial distribution of responses, and
hence, a different matrix of subjective distances. In what
follows, therefore, instead of analyzing the specific
characteristics of individual cells, the average behaviour is
studied.

\begin{figure}[htdf]
\begin{center}
\leavevmode \epsfysize=6truecm \epsfxsize=9truecm
\epsffile{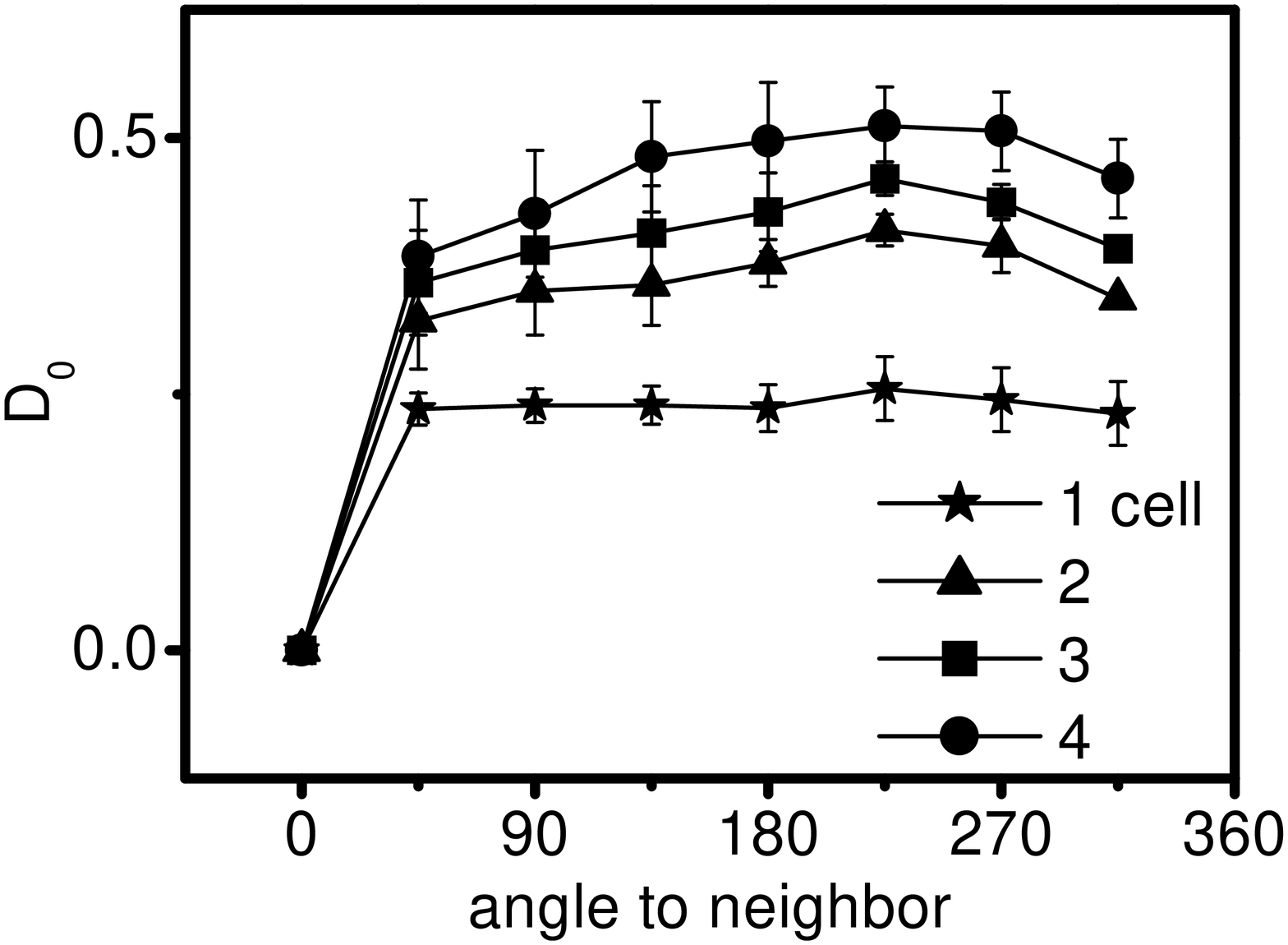}
\end{center}
\caption{Average distance of a given arm with all the others,
numbered counter-clockwise, as calculated from 62 units (29
hippocampal pyramidal neurons and 33 lateral septal cells). The
error bars are the standard deviations from the averages. Stars,
triangles, squares and circles correspond to 1, 2, 3 and 4 cells,
respectively.} \label{f5}
\end{figure}

In figure \ref{f5},
the average distance of a given arm with all the others is shown.
The numbers in the $x$ axis indicate the angle separating the two
arms under consideration. Thus, the average $D_0$ between all
pairs of arms which lie at 45$^\circ$ from one another is shown
as a single point in the plot. Each data point represents an
average among  cells (62 units have been considered: 29 pyramidal
cells in the hippocampus, and 33 lateral septal cells). The error
bars show the standard deviation of the cell average. In the
lower curve, the arm is decoded from the activity of a single
cell. As the curves rise, the decoding makes use of more cells
(from 1 to 4) recorded simultaneously.

The first thing that can be noticed, is that as the number of
neurons increases, all the subjective distances grow. This means
nothing but that the representation of the different arms becomes
more and more distinctive, and therefore, the fraction of
mistakes goes down. One can also observe that the single neuron
case looks pretty flat. That is, there is no evident structure in
the set of stimuli, since each arm has roughly the same
probability to be confounded with any other arm. An analysis of
variance (ANOVA) of the distances obtained for pairs of arms at
$45^\circ$ and $180^\circ$ shows that they are not significantly
different ($p > 0.3$).

In contrast, as the number of neurons increases, the curves begin
to bend, showing larger distances for pairs of arms that lie
farther apart. The ANOVA test shows that the distances obtained
for pairs of arms at 45$^\circ$ is significantly different from
the ones at $180^\circ$ ($p < 0.0001$). This means that if the
response of 4 neurons is considered, then it is more probable to
confound a given location with a nearby location, than with a
distal one. Topography, hence, seems to emerge from population
coding, not from the single cell response.

This example shows that the subjective distance is, when read
from 4 simultaneously recorded cells, a monotonic function of the
true distance. This is indicates that {\sl near} in the actual
world corresponds to {\sl near} in the subjective perception, and
{\sl far} in the actual world corresponds to {\sl far} in the
subjective perception. In this sense, one can ensure that there
is continuity, or topography. If the subjective distance were
strictly equal to the actual distance, however, the upper curve of
figure \ref{f5} would have a linear rise from $0^\circ$ to
$180^\circ$, and then a similar linear decrease from $180^\circ$
to $360^\circ$. The curved shape of the upper trace of figure
\ref{f5} indicates that the scale in the subjective
representations somehow shrinks as the actual stimulus moves
farther away. This, in turn, shows that the system is better
designed to make fine distinctions between nearby stimuli than
between distal ones. The single cell response, instead, makes no
distinction at all between near or far. It only recognizes whether
the two arms are the same or not. All these non trivial
characteristics of the coding properties of hippocampal and
septal cells have been visualized in terms of our definition of
the subjective distance.

\section{Summary}

The subjective distance, as introduced here, is a way of measuring
how differently stimuli are perceived. The distance between any
two elements may be interpreted in terms of the average
performance when trying to infer the actual stimulus, if only the
response of the subject is known. In such a performance, the
trial to trial variability of the responses to each stimulus is
as important as the mean responses.

It should be noticed that the probability of confusion does not
only depend on the characteristics of the two stimuli under
consideration, but also, on all the other stimuli in the set
${\cal S}$. Hence, the distance between two given items may vary,
when the remaining stimuli in $S$ are modified. So here, as well
as in many other information, or discrimination analyses, the
choice of the set of stimuli is a highly relevant (and sometimes
difficult) issue in itself, which should not be neglected.

The distances $D$ and $D_0$ may have several applications, for
example, it may be of interest to compare the subjective distance
with some other objective measure of dis-similarity. Here, as
shown in section \ref{placecells}, $D_0$ has proven useful to
show that an increasingly topographic encoding of spatial
location arises, as the number of cells grows. In this case, as
the population of neurons increases, the topography of real space
seems to emerge. There are other cases, though, where the
subjective perception of certain objects raises clear
differentiating bounds between stimuli that are actually near, in
the so called {\it physical} space. For example, it is known that
during the first year of life, the exposure of infants to their
mother tongue builds up a very particular way of perceiving
phonetic information. Two sounds that are physically very
similar, but correspond to two different phonemes in the child's
language, are easily discriminated. Yet, the experimentalist can
design two sound waves differing even more in their physical
characteristics, but that the infant cannot distinguish, simply
because they are not distinct building blocks (phonemes) in his or
her own language (see for example Kuhl 1994). Another example is
the perception of facial expression (Young {\it et al.} 1997),
where although all the experimentalist can continuously morph the
picture of a happy face into that of an angry one, human
observers have a tendency to categorize them into distinct
emotions (full happiness or full anger). Our subjective distance
would be a good way to quantify these effects.

Another possible application would be to use the subjective
distance as the input to a multidimensional scaling algorithm.
This would allow to place the stimuli in a finite dimensional
space, and to gain further geometrical intuition about them.

\section*{Acknowledgements}

We would like to thank Alessandro Treves for his very useful
suggestions. This work has been supported with a grant of
Fundaci\'on Antorchas, one of the Human Frontier Science Program
No. RG 01101998B, and a NIMH grant 58755.

\section*{References}

\begin{enumerate}

\item[-]
Amari, S. and Nagaoka, H.  (2000) {\it Methods of Information
Geometry} USA: Oxford University Press and American Mathematical
Society.

\item[-]
Cover, T. M. and Thomas, J. A. (1991) {\it Elements of
Information Theory}. Wiley: New York.

\item[-]
Cox, T. F. and Cox M. A. (2000)  {\it Multidimensional Scaling,
Second Edition} CRC: Chapman and Hall.

\item[-] Dayan, P. and Abbot L. F. (2001) {\it Theoretical Neuroscience: Computational and
Mathematical Modeling of Neural Systems}. The MIT Press: London.

\item[-] Green, D. M. and Swets J. A. (1966) {\it Signal Detection Theory and
Psychophysics}. Willey: New York.

\item[-] Kuhl, P. K (1994)Learning and representation in speech
and language. {\it Curr. Op. Neurobiol.} {\bf 4} 812 - 822

\item[-] Leutgeb, S. and Mizumori, S. J. (2002). Temporal
correlations between hippocampus and septum are controlled by
enviromental cues: evidence from parallel recordings. Submitted.

\item[-] Leutgeb, S. and Mizumori S. J. (2002) Context-specific spatial representations
by lateral septal cells. {\it J. Neurosci.} {\bf 112} (3) 655 -
663

\item[-] O'Keefe, J. and Dostrovsky, J. (1971). The hippocampus as a
spatial map: Preliminary evidence from unit activity in the
freely moving rat. {\it Brain Res.}, {\bf 34}, 171-175.

\item[-] O'Keefe, J. and Nadel L., (1978). {\it The hippocampus as a cognitive
map}. Oxford: Clarendon Press.

\item[-] Redish, A. D. (1999){\it Beyond the Cognitive map: From Place Cells to
Episodic Memory}. Cambridge, Massachusetts: The MIT Press.

\item[-] Rieke F., Warland D. de Ruyter van Steveninck R. and
Bialek W. (1997) {\it Spikes: exploring the neural code}.
Cambridge, Massachusetts: The MIT Press.

\item[-]
Rolls, E. T. and Treves, A. (1998) {\it Neural Networks and Brain
Function}. London: Oxford University Press.

\item[-]
Samengo, I. (2001) The information loss in an optimal maximum likelihood
decoding, {\it Neural Comput.}, {\bf 14} 771 - 779

\item[-]
Treves, A. (1997) On the perceptual structure of face space.{\it
Biosystems} {\bf 40} 189 - 196

\item[-] Victor, J. D. and Purpura K. P. (1997) Metric-space analysis of
spike trains: theory, algorithms and application. {\it Network: Comput.
Neural Syst.} {\bf 8} 127 - 164

\item[-]
Young F. W. and Hamer R. M. (1994) {\it Theory and Applications of
Multidimensional Scaling}. Hillsdale: Eribaum Associates.

\item[-]
Young, A.W., Rowland, D., Calder, A.J., Etcoff, N.L., Seth, A. and
Perrett, D.I. (1997). Facial expression megamix: tests of
dimensional and category accounts of emotion recognition. {\it
Cognition} {\bf 63} 271 - 313

\item[-] Zhou, T. L., Tamura R., Kuriwaki J. and Ono T.(1999) Comparison
of medial and lateral septal neuron activity during performance of spatial
tasks in rats. {\it Hippocampus} {\bf 9} 220 - 234

\end{enumerate}

\end{document}